\documentclass[sigconf]{acmart}

\def\BibTeX{{\rm B\kern-.05em{\sc i\kern-.025em b}\kern-.08emT\kern-.1667em\lower.7ex\hbox{E}\kern-.125emX}}

\copyrightyear{2019}
\acmYear{2019}
\setcopyright{acmlicensed}
\acmConference[]{}
\acmBooktitle{}

\usepackage{tikz}
\usepackage{amsmath}
\usepackage{multirow}
\usepackage{lipsum}
\usepackage{url}
\usepackage{todonotes}
\usepackage{booktabs}
\usepackage{makecell}
\usepackage{paralist}
\usepackage[binary-units=true]{siunitx}
\usepackage{csquotes}
\usepackage{enumitem}

\usepackage{caption}
\usepackage{subcaption}

\usepackage{pgfplots}

\usepackage{balance}

\usepgflibrary{patterns}

\pgfplotsset{
  grid style = {
    dash pattern = on 0.15mm off 1mm,
    line cap = round,
    gray,
    line width = 0.5pt
  }
}

\pgfplotsset{compat=1.14}

\newcommand{\coldbench}{\textsc{ColdBench}}

\begin{document}



\title{Cold Storage Data Archives: More Than Just a Bunch of Tapes}



\author{Bunjamin Memishi}
\affiliation{%
  \institution{German Aerospace Center}
  \city{Institute of Data Science}
  \country{Jena}
}
\email{bunjamin.memishi@dlr.de}

\author{Raja Appuswamy}
\affiliation{%
  \institution{EURECOM}
  \city{Biot}
  \country{France}
}
\email{raja.appuswamy@eurecom.fr}

\author{Marcus Paradies}
\orcid{0000-0002-5743-6580}
\affiliation{%
  \institution{German Aerospace Center}
  \city{Institute of Data Science}
  \country{Jena}
}
\email{marcus.paradies@dlr.de}

\begin{abstract}
The abundance of available sensor and derived data from large scientific experiments, such as earth observation programs, radio astronomy sky surveys, and high-energy physics already exceeds the storage hardware globally fabricated per year.
To that end, \emph{cold storage data archives} are the---often overlooked---spearheads of modern big data analytics in scientific, data-intensive application domains.
While high-performance data analytics has received much attention from the research community, the growing number of problems in designing and deploying cold storage archives has only received very little attention.


In this paper, we take the first step towards bridging this gap in knowledge by presenting an analysis of four real-world cold storage archives from three different application domains. In doing so, we highlight (i) workload characteristics that differentiate these archives from traditional, performance-sensitive data analytics, (ii) design trade-offs involved in building cold storage systems for these archives, and (iii) deployment trade-offs with respect to migration to the public cloud. Based on our analysis, we discuss several other important research challenges that need to be addressed by the data management community.

\end{abstract}

\maketitle

\section{Introduction}
\label{sec:introduction}

Data-intensive scientific domains are nowadays capable of generating large data volumes (\SI{}{\tera\byte}s to \SI{}{\peta\byte}s) within short time frames.
The data often stems from observational sciences, such as, earth observation, radio astronomy, nuclear physics, and medicine.
The immense scientific value of this data often lies in the ability to capture changes of the observed system over time and in enabling researchers to reason about the root cause of such changes.
To facilitate such time series analysis, scientific data is stored in \emph{cold storage data archives}, which serves two main purposes:
\begin{inparaenum}[(1)]
\item ensuring long-term data preservation and
\item providing a data platform for data analysis at large scale covering the complete project/experiment.
\end{inparaenum}

\noindent
The development of new Exascale supercomputing facilities has resulted in a dramatic increase in the capacity of these cold storage archives. Thus, in order to keep the total cost of ownership low, data was primarily stored in nearline storage systems, such as, tape libraries. However, sparked by increasing storage capacity demands, there is a growing interest in the development of more durable and high density storage media, including archival disks~\cite{Balakrishnan2014,Black2016}, DNA~\cite{Bornholt2016,Appuswamy19}, and optical media~\cite{Yan2017,Anderson2018} for cold storage. All these technologies vary dramatically with respect to read/write latency, available bandwidth capacity, and media durability. 
In order to systematically analyze the applicability of these new technologies for archiving scientific data, one needs a benchmarking framework that takes into account various requirements from the application domain to identify the optical set of storage devices. Unfortunately, no such cold storage benchmark exists today. 

With the increasing capacity of cold storage archives, configuration and tuning have become tedious and labor-intensive tasks. The computationally-intensive statistical techniques that are used to analyze scientific data also makes resource allocation and performance isolation a complex problem at multi-mission data archives that potentially span dozens of projects and experiments with vastly different storage and access requirements. 
Recently, several cloud service providers have started offering fully-managed, elastic, cold-storage-as-a-service platforms~\cite{url:glacier,url:google,url:microsoft} that solve some of these problems. However, little attention has been paid to understanding the advantages and disadvantages involved in migrating scientific data archives to the public cloud.

Scientific application domains also differ widely with respect to their data access demands from cold storage archives. Some domains require the cold storage to behave as an \emph{active archive}, where all data must be online and available at any point in time, as data retrieval of individual files or batches of files is common.
This holds true in particular for application domains that need to provide a consistent view across data gatherings spanning multiple decades of observations.
On the contrary, other domains require cold storage to act as a \emph{static archive} where most data is never read back again and only stored for long-term preservation purposes. The choice of media used for provisioning scientific data archival obviously depends on the nature of the archive. While prior studies have explored some characteristics of static archives, there have been very few studies on understanding data access patterns and deployment scenarios (in-house or cloud) for active archives.




\begin{figure*}

\begin{minipage}{0.6\textwidth}
\captionof{table}{Dataset characteristics.}
\vspace{-10pt}
\label{tbl:application-domain-characteristics}
\begin{center}
	\begin{tabular}{ r r  r  r  r}
\toprule
    & \textbf{ECFS}  &   \textbf{MARS}    &   \textbf{D-SDA}   &   \textbf{LOFAR} \\
	\midrule
Total capacity  &   \SI{14.8}{\peta\byte} &  \SI{37.9}{\peta\byte} & \SI{15}{\peta\byte}  & \SI{13}{\peta\byte}  \\
Ann. Growth   &   3+ PB/y   &   15+ PB/y & 10+ PB/y  &   2+ PB/y \\
Cache   & \SI{340}{\tera\byte}  & \SI{1}{\peta\byte}  &    \SI{500}{\tera\byte} & \SI{740}{\tera\byte}  \\
\#Files   & $137.5$ mil.  & $9.7$ mil.  & $128+$ mil. & $3.2+$ mil.  \\
Avg. file size   & $<1$ GB  & $1-64$ MB  & \SI{59.8}{\mega\byte}  & $<1$ GB  \\
Largest file    & \SI{32}{\giga\byte}  & \SI{1.34}{\tera\byte}  & \SI{38.49}{\giga\byte}  & \SI{63}{\giga\byte}  \\
\bottomrule
    \end{tabular}
\end{center}
\end{minipage}%
\hfill
\begin{minipage}{0.4\textwidth}
\captionof{table}{Workload characteristics.}
\vspace{-10pt}
\label{tbl:application-domain-workload-characteristics}
\begin{center}
\begin{tabular}{ r r r }
\toprule
& \textbf{ECFS} & \textbf{MARS} \\
\midrule
Never read files number   & $101.3$ mil.  & $7.9$ mil.  \\
Never read files size   & \SI{11.3}{\peta\byte} & \SI{24.9}{\peta\byte} \\
Never read files percentage   & $76\%$ & $80\%$ \\
\bottomrule
\end{tabular}
\end{center}
\end{minipage}
\end{figure*}

\label{sec:1.2-introduction}

Analyzing scientific data archives is currently not in the focus of commercial storage system providers and data management researchers.
We argue that there is an increasing need for such an analysis as it supports system designers to evaluate crucial architecture decisions for various aspects of data management (e.g., cache sizing \& eviction strategy, data placement, data prefetching strategies, etc.) for large-scale scientific projects.
Such an analysis also helps to ponder the performance--price trade-offs between private storage infrastructures, hybrid private/public cloud storage, and public cloud storage solutions for a data-intensive scientific project.
Finally, such an analysis can also drive the development of a benchmark designed for evaluating cold storage data archive systems independent of the specific application domain. 

In this paper, we take a first step towards bridging this gap by performing an analysis of four real-world data archives with a focus on the domain-specific characteristics of the corresponding storage systems to (i) highlight key workload characteristics that differentiate scientific data storage from traditional, performance-sensitive data storage, (ii) describe design trade-offs involved in building cold data storage systems tailored for scientific data archival, and (iii) explore deployment trade-offs with respect to migration to the public cloud. Based on our analysis, we discuss open challenges in the area of cold storage data archives to be tackled by the research community.


\noindent


\section{Application Domains}
\label{sec:applications}

In this section we describe three different scientific application domains, namely earth observation (D-SDA), radio astronomy (LOFAR), and weather forecasting (ECMWF), and detail specifics about their employed storage systems, data storage characteristics, and data access characteristics.

\medskip

\noindent\textbf{D-SDA:}
The D-SDA is operated by the Earth Observation Center (EOC) of the German Aerospace Center and is a multi-mission data management and information system covering national and international earth observation missions~\cite{2016:D:SDA:Kiemle}.
The data (and derived data) is stored in a large, geo-replicated cold storage data archive facility, which relies on a robotic tape library system.
Commonly used file formats in the D-SDA include image file formats, such as GeoTIFF and JPEG, but also scientific file formats, such as, netCDF and HDF5.
The accompanying metadata is stored in a relational database system and serves as identification \& localization service for end users.

Depending on the specific EO mission, various value-added data products can be generated upon arrival of the raw data at the ground station and are archived for later reuse.
Currently, most value-added data products are generated automatically using complex software pipelines and stored for faster retrieval later on.
Depending on mission-specific service-level agreements, value-added data products have to be available for public download within a specific, fixed time frame from the sensing timestamp (typically a few hours after sensing).
Besides the data-driven generation of value-added data products, data retrieval of any data item can be triggered at any time by users or by so-called reprocessing campaigns.
A reprocessing campaign often runs over multiple months and re-generates derived data products, when a new algorithm version or configuration becomes available.
Thus, data items have to be accessible at any point in time and render the D-SDA as a paramount example of an \emph{active archive}.

\medskip

\noindent\textbf{LOFAR:}
The LOFAR (\underline{LO}w \underline{F}requency \underline{AR}ray) radio telescope consists of a large array of individual antennas distributed across Europe.
These antennas form a single large, virtual radio telescope with a huge diameter of hundreds of kilometers.
During observation, the individual antenna signals are correlated and stored in the LOFAR long-term archive in the binary, astronomy-specific MeasurementSet file format~\cite{VanDiepen15}.
The LOFAR long-term archive is geo-distributed across multiple facilities in Europe, with the storage system at the J{\"u}lich Supercomputing Centre being the largest one.

An important application is the generation of celestial maps, where an iterative process transforms the received radio wave signals into viewable images.
Once the celestial map has been generated, the raw data typically remains in the long-term archive and is only rarely accessed.
Thus, according to our terminology, the LOFAR data archive is a \emph{static archive} since most (raw) data is never accessed again.

\medskip

\noindent\textbf{ECMWF:}
The weather forecasting storage system is represented with The European Centre for Medium-Range Weather Forecasts (ECMWF).
ECMWF~\cite{Grawinkel2015} produces global numerical weather predictions for its member states and a broader community.
Up to the time of writing, ECMWF operates one of the largest supercomputer facilities and data archives worldwide. ECMWF uses two archival systems that were developed in house, namely, ECFS, a general-purpose file archive that is used for long-term data storage, and MARS, a large-object database that stores meteorological data. Unlike ECFS, where data is stored as opaque files that are rarely accessed, MARS is a database that records domain-specific fields and exposes them to users using a customized query language. As users can access and retrieve any field at any time, MARS is an active archive compared to ECFS which is static in nature.

\section{Analysis}
\label{sec:zoom-out}

In this section, we present an analysis of the four archives described in Section~\ref{sec:applications} that span three application domains. We first present a data analysis in Section~\ref{sec:performance} to highlight unique properties of scientific data archives and their workloads. Then, we present a deployment analysis in Section~\ref{sec:deployment} to understand the pros and cons of using public, cloud-based, cold storage services for archiving scientific data.

\subsection{Data Analysis}
\label{sec:performance}

Table~\ref{tbl:application-domain-characteristics} shows various characteristics of the four archives used in our study.
The numbers reported for D-SDA and LOFAR are based on an in-house analysis we conducted on these archives.
For ECFS and MARS, the values reported are based on a previous study~\cite{Grawinkel2015}.

\begin{table*}[tb]
\caption{Commercial offerings.}
\vspace{-10pt}
\label{tbl:commercial-storage-offerings}
\begin{center}
	\begin{tabular}{ r r r r r }
	\toprule
& \makecell{Storage \\ (GB/Month)} & \makecell{Retrieval \\ (per GB)} & \makecell{GET Requests \\ (per 10,000)} & Latency \\
\midrule
Azure Archival Blob & \$$0.0045$ & \$$0.02$ & \$$0.5$ & several hours \\

Azure Cool Blob & \$$0.0334$ & \$$0.01$ & \$$0.01$ & \SI{61.4}{\milli\second} \\

Azure Hot Blob & \$$0.0422$ & \$$0$ & \$$0.004$ & \SI{5.3}{\milli\second} \\ 
\bottomrule
    \end{tabular}
\end{center}
\end{table*}



\smallskip

\noindent\textbf{Data volume.} The amount of data stored across all archives is in the order of tens of \SI{}{\peta\byte}s stored across millions of files.
As new exascale supercomputing technologies are deployed for scientific analysis, the amount of data stored by scientific archives continues to grow rapidly. As shown in Table~\ref{tbl:application-domain-characteristics}, these archives exhibit a 15\% to 65\% cumulative annual growth rate as they continue to add several Petabytes of data to their archival storage.
This rate of data growth is unsustainable in the long run, as several studies have pointed out that areal density improvements in available storage is far below this rate of data growth (16\% improvement in density per year for HDDs, and 33\% for tape)~\cite{horison-storage-outlook,Appuswamy2017}.
While researchers are investigating the feasibility of novel storage media, like DNA~\cite{Bornholt2016,Appuswamy19} or optical~\cite{Yan2017,Anderson2018} storage, for dramatically improving density, scientific archives will have little option but to implement means to reduce data growth for the foreseeable future.

\smallskip

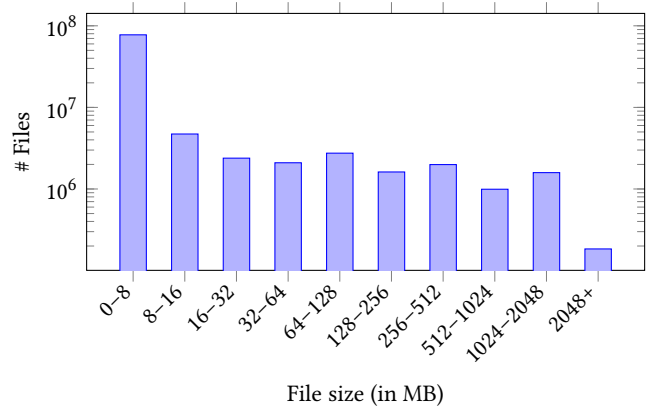
\begin{figure}[tb]
\centering
\begin{tikzpicture}
    \begin{semilogyaxis}[
            width=9cm,
            height=5cm,
    		ylabel={\# Files},
			xlabel={File size (in \SI{}{\mega\byte})},
            ybar,
            ymin=0,
            symbolic x coords={0--8, 8--16, 16--32, 32--64, 64--128, 128--256, 256--512, 512--1024, 1024--2048, 2048+},
            xtick=data,
            x tick label style={rotate=45,anchor=east}
        ]
        \addplot table[x=MB,y=NrFiles]{\filesizedistribution};
    \end{semilogyaxis}
\end{tikzpicture}
\vspace{-15pt}
\caption{File size distribution: Main D-SDA product library.}
\label{fig:file-size-distribution-d-sda}
\vspace{-10pt}
\end{figure}

\noindent\textbf{Data variety.} Considering the fact that these storage systems are applied to specific applications domains, there is another factor that should be taken into account, namely the data variety. Given that the latency of accessing data on cold storage devices can be quite high, one aspect of variety that is particularly important is the distribution of file sizes. We use D-SDA as an example to explore this. Table~\ref{tbl:application-domain-characteristics} shows that the average file size of D-SDA is around \SI{64}{\mega\byte}. Figure~\ref{fig:file-size-distribution-d-sda} shows the file size distribution of the main D-SDA product library, which hosts all EO products of the national multi-mission ground segment archived in Oberpfaffenhofen. Clearly, there is a huge variety in the sizes of files which are being saved. The DFD storage system is mostly used for storing data from different earth observation missions. Thus, starting from a file with a couple of kilobytes for specific observation parameters, the file size could easily reach a couple of gigabytes, and even more.

Given the prevalence of small files, several of the files reaccessed from tape are likely to be small in size. Small file retrieval is an inherently suboptimal access pattern for tape archives, as it leads to long-latency tape load/unload operations caused by random accesses. This is the reason why all scientific storage systems use a HDD-based caching layer to buffer all small files, and frequently accessed files, within their cache capacity. The caching layer also doubles in role as a burst buffer to temporarily stage new data before it is eventually moved to the tape backend. The actual ratio of data staged in HDD-based caches versus tape varies from 1:17 for LOFAR to 1:30 for D-SDA.
Depending on the specific requirements, caching ratios vary (i.e., due to budget limitations) and data archives either separate between read/write caching (caches for reads, burst buffers for writes) or only utilize a common caching layer for both.

\smallskip

\noindent\textbf{Data liveliness.} Another property of data that is common across all archives is the fact that a large fraction of data stored is rarely read again. Table~\ref{tbl:application-domain-workload-characteristics} shows the ``liveliness'' of data for ECFS and MARS. As can be seen, only 20\% of data in both archives is accessed after being stored.
Given that these file accesses are not performance critical, an ideal media for archival data should focus on optimizing the cost of long-term storage.
Tape offers the highest density, and the lowest cost/GB, among storage media available today.
Further, tape consumes no power once unmounted, and also has the longest media lifetime.
Due to these reasons, all four scientific archives rely on a data tape facility for long-term data storage.

\pgfplotstableread[row sep=\\,col sep=&]{
    MB & NrFiles \\
    0--8 & 77540744 \\
    8--16 & 4719466 \\
    16--32 & 2387125 \\
    32--64 & 2095864 \\
    64--128 & 2748315 \\
    128--256 & 1616620 \\
    256--512 & 1991281 \\
    512--1024 & 993066 \\
    1024--2048 & 1586496 \\
    2048+ & 184138 \\
    }\filesizedistribution

\subsection{Deployment Analysis}
\label{sec:deployment}

\noindent\textbf{On premise versus cloud.} Several cloud service providers have started offering archival storage as an elastic service. Thus, we will now explore the trade-offs involved in using a cloud-based cold storage service for archiving scientific data. Table~\ref{tbl:commercial-storage-offerings} presents price--performance metrics of popular archival-as-a-service offering from the Microsoft Azure cloud. Similar to Azure, all cloud providers offer three types of storage classes for archival storage with price--performance characteristics matching the expected workload. The first service in Table~\ref{tbl:commercial-storage-offerings}, Archival Blob store, is a \emph{Deep archival} storage service tailored towards storage of data that is very rarely accessed. The second service, Cool blob store, is a \emph{Nearline archival} storage service tailored for storing data that is more frequently accessed, but infrequent enough that storing it on non-archival services would incur additional expenses. Finally, Hot blob store is an \emph{online} service used for storing frequently accessed data.

The cost of storing data (second column) drops by an order of magnitude if one uses a deep archival service compared to an online service (cf.~Table~\ref{tbl:commercial-storage-offerings}).
Given that 80\% of data stored in static archives is never read back, deep archival in the cloud might be a good fit for such data. 
Similarly, nearline services often provide a 30\%--50\% reduction in storage cost compared to online storage.
Thus, the data stored in the HDD cache might be a good fit for these nearline services.
While these price points appear to make archival storage services an attractive option compared to on-premise storage for static archives, there is an important storage--access trade-off that must be considered before migrating to the cloud.

\smallskip

\noindent\textbf{Storage--access tradeoff.} When data is stored in online services, it is typically available for instant access. However, data stored in nearline or deep archival services need to be \emph{rehydrated} and temporarily staged in an online service before it can be accessed by an application. As a result, nearline and deep archival services charge both a rehydration fee and a data access fee, while online services only charge for data accesses. Thus, looking at Table~\ref{tbl:commercial-storage-offerings}, one can see that cost of retrieving data from the storage service follows the inverse path of raw storage cost--while storage cost increases as one moves from deep archival to online storage, data retrieval cost decreases an order of magnitude in the same direction. This inverse relationship between storage and retrieval cost has important implications on the deploying scientific archives in the public cloud.

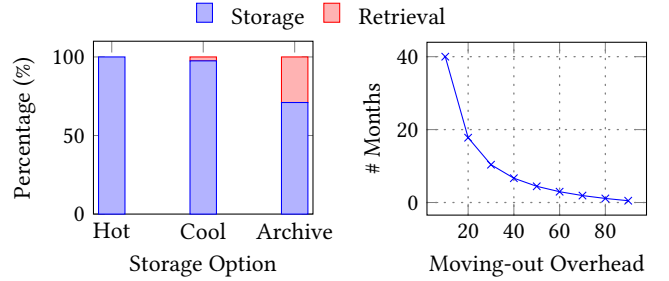
\begin{figure}[t]
\ref{costbreakdown}\\
\centering
\begin{subfigure}[t]{0.45\columnwidth}
\begin{tikzpicture}%
\begin{axis}[width=4.5cm,
             ymin=0,
			 ylabel={Percentage (\%)},
			 ybar stacked,
			 xtick=data,
			 xlabel={Storage Option},
			 symbolic x coords={Hot, Cool, Archive},
			 legend style={at={(0.7,0.75)},anchor=north,draw=none,column sep=5pt},legend columns=-1,legend to name=costbreakdown]
\addplot coordinates {(Hot,100) (Cool,97.5) (Archive,71)};
\addplot coordinates {(Hot,0) (Cool,2.5) (Archive,29)};
\legend{Storage,Retrieval}
\end{axis}
\end{tikzpicture}
\caption{Cost breakdown.}
\label{fig:cost-breakdown}
\end{subfigure}
\hfill
\begin{subfigure}[t]{0.45\columnwidth}
\begin{tikzpicture}
\begin{axis}[width=4.5cm,
             xlabel={Moving-out Overhead},
             ylabel={\# Months},
             ymajorgrids,
             xmajorgrids]
\addplot[color=blue,mark=x] coordinates {
	(10,40)
	(20,17.7778)
	(30,10.37037)
	(40,6.66667)
	(50,4.4444)
	(60,2.96296)
	(70,1.904)
	(80,1.11111)
	(90,0.4938)
};
\end{axis}
\end{tikzpicture}
\caption{Storage--access trade-off.}
\label{fig:storage-access-relationship}
\end{subfigure}
\caption{Cloud migration analysis.}
\label{fig:overall_costs}
\end{figure}

The choice of storage between the nearline and archival tier very much depends on the archival workload. As an example, let us consider a 1PB scientific archive that is stored for a year and read back in its entirety just once during the entire year. Based on pricing details shown in Table~\ref{tbl:commercial-storage-offerings}, assuming a blob size of 256MB, the overall cost of the archive would be \$79K, \$430K, and \$531K for the archival, cool, and hot blob storage services, respectively. Figure~\ref{fig:cost-breakdown} shows the relative breakdown of the cost to separate out the contribution of raw storage and data accesses. As can be seen, there is a huge difference between archival blob store and the rest in that 30\% of the overall costs can be attributed to reading back data in the former case. The per-GB data retrieval cost charged for archival storage is the dominating source of this 30\%. Thus, using the simple cost equation

\begin{equation*}
\setlength{\belowdisplayskip}{5pt} \setlength{\belowdisplayshortskip}{5pt}
\setlength{\abovedisplayskip}{-5pt} \setlength{\abovedisplayshortskip}{-5pt}
\begin{aligned}
    TotalCost = StorageCost \times M + ReadCost \times R
\end{aligned}
\end{equation*}

\noindent
where M is the number of months and R is the number of times data is retrieved completed, we can derive the overhead of data access if data is accessed every month ($M = R$). Using pricing information from Table~\ref{tbl:commercial-storage-offerings}, we compute it to be 82\%. 
If we assume we access data once a year for $R$ years, then, M is $R \times 12$. Based on pricing numbers in Table~\ref{tbl:commercial-storage-offerings}, the overhead of scanning data once a year is 27\%. These results indicate that cloud storage might be more suited for static archives with little to no data access. Active archives, in contrast, need much more frequent access to data. Thus, migrating active archives to the cloud will lead to storage no longer being the dominating cost, which is ironic given that storage cost the main motivating factor behind cloud migration of these archives. 

\smallskip

\noindent\textbf{Data scrubbing and vendor lock-in.} The aforementioned storage--access trade-off presents two additional problems even for static archives, namely data scrubbing and vendor lock-in. First, all static archives routinely scrub data to ensure data integrity and to protect data from corruption due to media failures. Our analysis indicates that scrubbing can be an expensive proposition in cloud-based static archives. As we mentioned earlier, the overhead of accessing data once a year in the cloud is 27\% in our scenario. Note that this does not include the network utilization charges for transferring data between the storage and compute nodes, which is quoted separately by all cloud service providers. Unless cloud-service providers offer built-in data scrubbing as a part of the service offering, data verification costs will be a non-negligible amount of the overall expenditure. 

\noindent
Second, once a data archive has been migrated to the cloud, moving back out of the cloud requires accessing all data once. We can use the former equation to derive the number of months data should be stored for this one-time, moving-out overhead to be a small fraction of the total cost. Figure~\ref{fig:storage-access-relationship} plots the relationship between the number of months and this overhead. As can be seen, in order for the moving-out overhead to be less than 10\% of the total cost, data must be stored for at least 40 months. Viewed another way, the cost of migrating 1PB of data out of the cloud (\$23K) is equivalent to storing it in the cloud for an additional 5 months based on cost metrics given in Table~\ref{tbl:commercial-storage-offerings}. Note that this cost does not include egress charges out of the cloud which are billed separately. For instance, the lowest egress charges from Azure are \$0.05/GB for outbound transfers. Including this would make total migration charge of \$75K for 1PB, which is equivalent to 16 months of storage. This clearly indicates that once a scientific archive is migrated to the public cloud, the economic incentive for moving out is very low. Given that the storage pricing across cloud providers is similar, the incentive for moving to another cloud is even lower due to the additional data ingestion charges that have to be paid.


\smallskip

\noindent\textbf{Tiered cold storage archive.} Based on our analysis, a two-tier, hybrid cloud infrastructure seems to be more appropriate for scientific data archives. Such an approach would store one copy of archival data locally and one or more copies in the cloud. This setup would solve several problems that complicate migration of scientific archives to the cloud. First, if all data access operations can be limited to the local copy, this approach would eliminate the associated cloud data retrieval overheads. Second, data scrubbing can be done on the local copies, and the cloud copies can serve as backup in case of local failures. In fact, one could improve availability by storing copies across multiple cloud service providers. Third, the local copy would solve the problem of vendor lock-in as it no longer needs to retrieve back the data during cloud migration.

\section{Discussion}
\label{sec:discussion}

The analysis and observations made in Section~\ref{sec:zoom-out} provide an interesting starting point for furthering our understanding of the challenges involved in designing, operating, and monitoring cold storage data archives for data-intensive scientific domains.
In this section, we discuss other important open problems and interesting research challenges that, we believe, require further attention.


\subsection{Active Archive---A Tertiary Polystore}
\label{sec:metadata}

All the four archives considered in this study store hundreds of millions of files, which store domain-specific, structured data using optimized file formats. The hierarchical organization of files, the mapping of domain-specific entities to files, assignment of files to tape drives, and auxiliary metadata generated by data mining tasks that crawl the archive are all stored and managed separately. For instance, CERN's Tape Archive system~\cite{Murray_2017}, which provides the tape-backend for storing data from Large Hadron Collider experiments stores its file catalogue in a relational database while storing the data itself in an object store. D-SDA stores accompanying metadata that serves as identification and localization service for end users in a relational database system.

In contrast to the physical organization of files on tape media, data mining tasks and computational models often work with domain-specific representations of this data. Thus, these archives also provide customized query languages to enable search and retrieval functionality at a ``logical'' level. For instance, MARS hosts 170 billion fields of meteorological data in 9.7 million files. Users do not directly access the fields, but issue a query using a custom query language. Thus, it is important for active archives to support access methods that can be used to answer user queries. 

Finally, unlike static archives where data stored is never accessed again, any data stored in these active archives can be requested at any point in time. While performance is not a priority, it is still important to apply scheduling techniques and caching hierarchies that are customized to the archive's workload in order to avoid pathological scenarios. For instance, MARS uses a separate Field DataBase (FDB) to cache fields that are frequently accessed. In addition, MARS also uses disk arrays as second-level file caches in front of tape drives. Despite the use of such deep caching hierarchies, and despite the fact that these caches have been reported to have a 95\% hit rate, the volume of accesses from 5\% of misses is high enough to heavily stress the tape robots. Thus, researchers have demonstrated the use of MARS-specific tape prefetching and request scheduling algorithms that improve performance.

These requirements of active archives make it a prime candidate for adopting a polystore architecture. However, while current work focuses on using polystores for performance-sensitive analytics, the use of polystores for managing cold data in scientific data archives presents interesting research challenge at the other end of the storage spectrum.
\subsection{Towards a Cold Storage Benchmark}
\label{sec:benchmarking}

In Section~\ref{sec:zoom-out}, we presented an in-depth exploration of several aspects of a cold storage system design.
In order to compare and trade off different design decisions, an independent, consistent, and comprehensive benchmark for cold storage systems and services is required. Such a benchmark does not only provide a convenient way to evaluate different available design options, but it also facilitates users to pose critical \emph{what-if} questions to adapt to changing application requirements and storage technology changes.

There are a few desirable properties that a cold storage benchmark for scientific applications should consist of.
It should be relevant for the application, i.e., it should resemble the real workload and data distribution accurately.
Most scientific workloads can be characterized as follows:
\begin{inparaenum}[(i)]
\item data is accessed only infrequently and
\item a limited number of users (power users) generates large parts of the read workload.
\end{inparaenum}
The benchmark should also be economical (preferably open and free), portable, extensible, and support private, hybrid, and public (cloud-based) cold storage systems.

\begin{figure}[tb]
\centering
\includegraphics[width=\columnwidth]{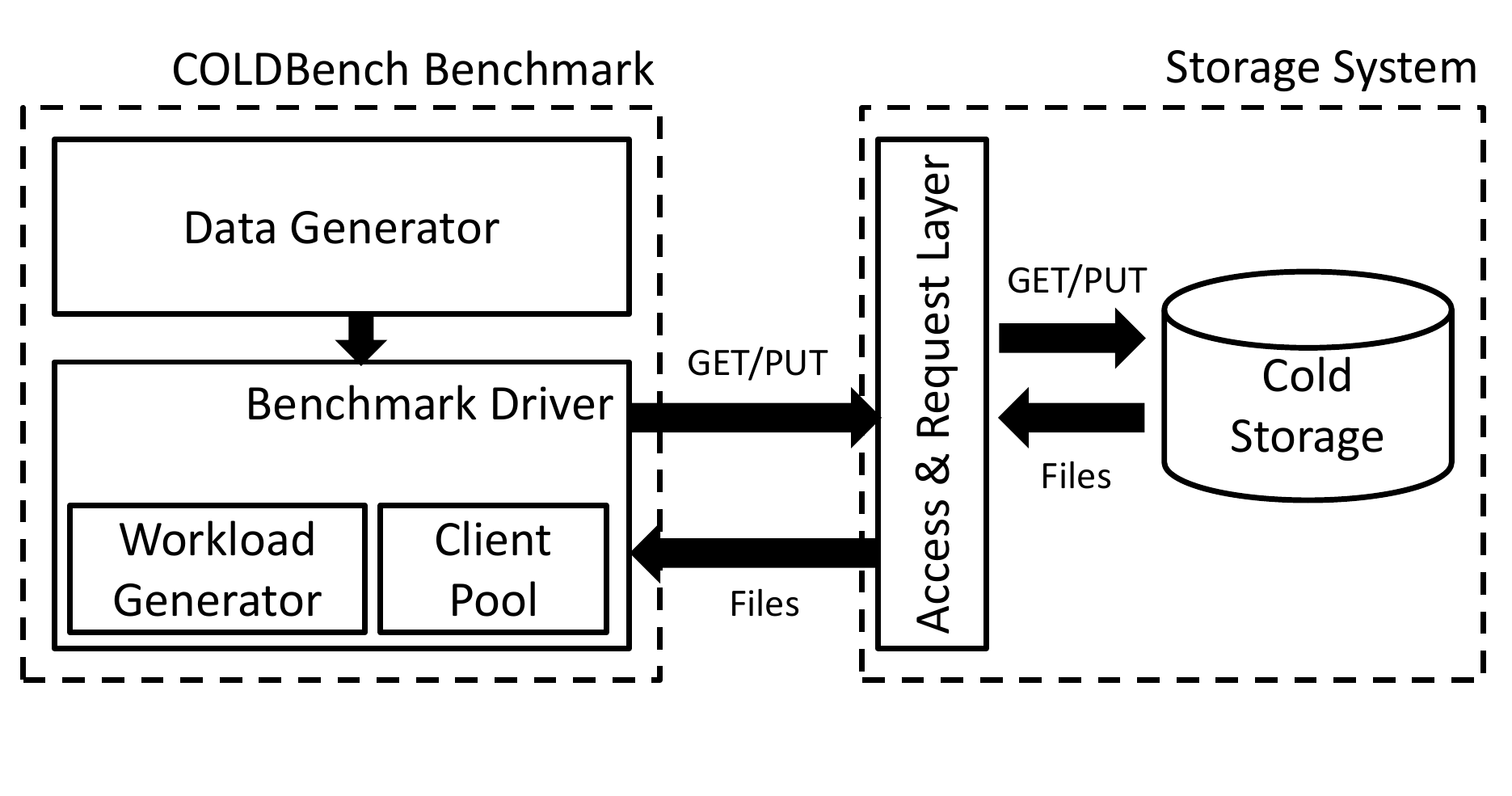}
\vspace{-35pt}
\caption{\coldbench{} architecture overview.}
\vspace{-10pt}
\label{fig:benchmark_architecture}
\end{figure}

Given these requirements, we are developing \coldbench{}, a benchmarking framework for evaluating and comparing heterogeneous cold storage systems.
We sketch the high-level architecture of \coldbench{} in Figure~\ref{fig:benchmark_architecture}.
It consists of two major components, namely the \emph{Data Generator} and the \emph{Benchmark Driver}, and connects to a cold storage system under test using a simple \texttt{GET/PUT}-like API. In the rest of this section, we describe the requirements that an ideal data generator and benchmark driver should meet, and some of the challenges involved in building these components.

\smallskip

\noindent\textbf{Data Generator.}
The data generator should generate files of varying size following a user-defined, application-specific file size distribution.
The specific data distribution should be derived from a real-world scenario or can be selected from a list of predefined, commonly observed file size distributions.
The file size distribution should be skewed, with outliers of extremely small files (\SI{}{\kilo\byte}s) to large files (\SI{}{\tera\byte}s).

The data generator should scale to generate data sets of up to multiple \SI{}{\peta\byte} without compromising the specifics of the file size distribution.
In order to mimic the write workload of a cold storage system, the data generator should produce a large, static data set, which gets populated initially and a smaller, dynamic data set, which gets added interweaved with the read workload.


\smallskip

\noindent\textbf{Benchmark Driver.}
The benchmark driver consists of a workload generator and a client pool. The workload generator should produce a sequence of read/write operations, which can be either single or batch requests.
The client pool allows instantiating a user-defined number of concurrent user sessions, which each run an individually generated workload against the cold storage system.
The workload generator should be configurable in the ratio of read/write operations, the ratio of single/batch requests, and the specification of domain-specific data priorities.
For example, in the D-SDA storage system, most of the user request workloads are focused on a particular earth observation mission~\cite{2019:BiDS:Schindler}.
Thus, representing temporal and spatial locality is a crucial aspect for the generation of realistic, domain-specific benchmark workloads.




\smallskip

\noindent\textbf{Choke Point-based Design.}
We propose a choke point-based benchmark design, which encompasses key technical challenges that real-world cold storage systems often face in operational settings~\cite{Boncz14}.
A choke point-based design ensures that the benchmark workload covers bottlenecks often observed in operational systems and forces cold data archive providers onto a path of continuous technological innovation.
Based on our analysis in Section~\ref{sec:zoom-out}, we can already derive initial choke points that a cold storage system has to deal with in practice. This includes
\begin{inparaenum}
 \item skewed data access,
 \item large batch file requests,
 \item dealing with different data retrieval priorities, and
 \item handling data access to small files efficiently.
\end{inparaenum}

\smallskip

\noindent\textbf{Benchmark Metrics.}
A cold storage benchmark should facilitate in addition to performance-related metrics, such as latency and sustained download bandwidth, also a cost-related metric.
This cost-related metric includes hardware costs (initial infrastructure and hardware replacement costs), administration \& utility costs, and potential software license costs.
This is in particular challenging for private cold storage systems, where the overall costs cannot always be easily derived.
In contrast, public, cloud-based cold storage services offer fine-grained billing and online calculators for anticipated costs considering the data set characteristics and the workload are known in advance~\cite{url:glacier,url:google,url:microsoft}.



\subsection{Provisioning \& Configuring Data Archives}
\label{sec:provisioning}

Storage system provisioning, configuration, and setup is becoming increasingly complex and tedious attributable to the thriving, ever-growing number of offerings by public cloud providers and hardware manufacturers.
A customer can choose between various private, hybrid, and public storage system offerings with greatly varying performance and cost characteristics.

For private storage infrastructures, storage hardware options are becoming increasingly multifarious---there are multiple storage media to choose from, e.g., HDDs (in installations of massive arrays of idle disks), tapes, flash-based storage, or optical storage systems.
Further, modern computer networks exhibit large bandwidths, low latency, and programmability of the network devices (e.g. smart switches and NICs).
Depending on the specific target application and its data- and workload characteristics, vastly different system provisioning and configuration considerations have to be taken into account.
This in turn requires tools to simulate and evaluate different system configurations in a comprehensive manner (cf.~Section~\ref{sec:benchmarking}), advisory tools that assist customers to select the best performing and cost-efficient system configuration, and full-system monitoring (cf.~Section~\ref{sec:monitoring}).

\subsection{Archive Profiling \& Monitoring}
\label{sec:monitoring}

Current data archives, with particular emphasis on the private storage systems, are still lacking  extensive evaluation and analysis, which is conditioned in having an appropriate monitoring and tracing system.
An end-to-end monitoring system would simplify a storage system evaluation and its improvement on the performance and the reliability context, among others.

\noindent
From our own experience, the current methodologies for gaining knowledge about the internals of private data archives have been inappropriate and time-consuming.
In the case of D-SDA and LOFAR, getting access to the cold storage data archive traces was conditioned by many obstacles, mainly based on twofold reasons:
\begin{inparaenum}[(1)]
\item complexity and
\item privacy.
\end{inparaenum}
Every storage system layer was having a proper tracing methodology, and the logical matching of the data archive events was not very straightforward.
The other side of the coin (that is, the privacy), implied every trace request to be followed by a lengthy period of weeks and months, until getting a permission for analyzing the particular storage system layer traces, even for trace data which was assumed to be open and free.

In the same time of designing a cold storage data archive, the application domain leaders should concurrently explore different monitoring systems, which could potentially be used as a fundamental framework in tracing their data archive.
After an extensive evaluation of the proposals, they will have to choose, modify or come up with a reasonable alternative, which encapsulates a certain number of modules that hide the inter-layer complexity and enable a customizable privacy, on top of an existing or new prototypical monitoring system.
If complexity requires the understanding of different storage system layers and their intersections, the privacy issue should clearly define the boundaries of what sensitive data is and what is not.
An end-to-end monitoring and tracing storage system should be capable of giving an efficient and real-time pipeline view at the granularity of individual user request/response operations.
In this way, even if leaded from an intuition~\cite{2017:SOSP:Canopy:Facebook}, one could accelerate an analysis and solution of a probable bottleneck, such as the tail latency.

\vspace{-2pt}
\section{Summary}
\label{sec:summary}

In this paper we make the case for cold storage data archives as fundamental building blocks for data-intensive, scientific application domains, such as, earth observation, radio astronomy, and weather forecasting. Consequently, we took the first step towards understanding the challenges involved in scientific data archival. Using a detailed analysis of four real-world, scientific, cold storage data archives, we demonstrated the heterogeneous nature of application workloads and showed that a hybrid two-tier approach with a combination of a private and a public cold storage infrastructure is most promising for a reasonable cost/performance trade-off.

We believe that cold storage data archives are largely overlooked by the research community although they entail a variety of interesting and challenging research questions. We discussed several such areas of exploration to highlight the fact that scientific data archival is not just a storage problem, but a rich data management problem with research challenges that span all important steps in the data management life cycle, ranging from planning and provisioning, performance monitoring \& tuning to keep the storage system in a healthy state, to providing a seamless view across metadata and experimental data to the end user through a common data management abstraction with querying/analysis capabilities.
Finally, we envision that cold storage data archives, in particular active archives, will exhibit an increased interest both from industry and the research community, due to storage specializations towards vastly different deployment areas and recent advances in storage hardware development.



\balance

\clearpage

\bibliographystyle{plain}
\bibliography{references}

\begin{thebibliography}{10}

\bibitem{url:glacier}
{\textsc{Amazon Glacier}}.
\newblock \url{https://aws.amazon.com/de/glacier/}.
\newblock Accessed: 01-02-2019.

\bibitem{url:google}
{\textsc{Google Archival Cloud Storage}}.
\newblock \url{https://cloud.google.com/storage/archival/}.
\newblock Accessed: 01-02-2019.

\bibitem{url:microsoft}
{\textsc{Microsoft Cool Blob Storage}}.
\newblock
  \url{https://azure.microsoft.com/en-us/blog/introducing-azure-cool-storage/}.
\newblock Accessed: 01-02-2019.

\bibitem{Anderson2018}
Patrick Anderson, Richard Black, Ausra Cerkauskaite, Andromachi
  Chatzieleftheriou, James Clegg, Chris Dainty, Raluca Diaconu, Rokas
  Drevinskas, Austin Donnelly, Alexander~L. Gaunt, Andreas Georgiou,
  Ariel~Gomez Diaz, Peter~G. Kazansky, David Lara, Sergey Legtchenko, Sebastian
  Nowozin, Aaron Ogus, Douglas Phillips, Antony Rowstron, Masaaki Sakakura,
  Ioan Stefanovici, Benn Thomsen, Lei Wang, Hugh Williams, and Mengyang Yang.
\newblock {Glass: A New Media for a New Era?}
\newblock In {\em 10th {USENIX} Workshop on Hot Topics in Storage and File
  Systems (HotStorage 18)}, Boston, MA, 2018. {USENIX} Association.

\bibitem{Appuswamy2017}
Raja Appuswamy, Renata Borovica{-}Gajic, Goetz Graefe, and Anastasia Ailamaki.
\newblock {The Five-minute Rule Thirty Years Later and its Impact on the
  Storage Hierarchy}.
\newblock In {\em International Workshop on Accelerating Analytics and Data
  Management Systems Using Modern Processor and Storage Architectures,
  ADMS@VLDB 2017, Munich, Germany, September 1, 2017.}, pages 1--8, 2017.

\bibitem{Appuswamy19}
{R}aja {A}ppuswamy, {K}evin {L}ebrigand, {P}ascal {B}arbry, {M}arc {A}ntonini,
  {O}livier {M}adderson, {P}aul {F}reemont, {J}ames {M}c{D}onald, and {T}homas
  {H}einis.
\newblock {O}ligo{A}rchive: {U}sing {DNA} in the {DBMS} storage hierarchy.
\newblock In {\em {B}iennal {C}onference on {I}nnovative {D}ata {S}ystems
  {R}esearch}, CIDR '19, 2019.

\bibitem{Balakrishnan2014}
Shobana Balakrishnan, Richard Black, Austin Donnelly, Paul England, Adam Glass,
  Dave Harper, Sergey Legtchenko, Aaron Ogus, Eric Peterson, and Antony
  Rowstron.
\newblock {Pelican: A Building Block for Exascale Cold Data Storage}.
\newblock In {\em 11th {USENIX} Symposium on Operating Systems Design and
  Implementation ({OSDI} 14)}, pages 351--365, Broomfield, CO, 2014. {USENIX}
  Association.

\bibitem{Black2016}
Richard Black, Austin Donnelly, Dave Harper, Aaron Ogus, and Anthony Rowstron.
\newblock {Feeding the Pelican: Using Archival Hard Drives for Cold Storage
  Racks}.
\newblock In {\em 8th {USENIX} Workshop on Hot Topics in Storage and File
  Systems (HotStorage 16)}, Denver, CO, 2016. {USENIX} Association.

\bibitem{Boncz14}
Peter Boncz, Thomas Neumann, and Orri Erling.
\newblock {TPC-H Analyzed: Hidden Messages and Lessons Learned from an
  Influential Benchmark}.
\newblock In {\em Performance Characterization and Benchmarking}, pages 61--76.
  Springer International Publishing, 2014.

\bibitem{Bornholt2016}
James Bornholt, Randolph Lopez, Douglas~M. Carmean, Luis Ceze, Georg Seelig,
  and Karin Strauss.
\newblock {A DNA-Based Archival Storage System}.
\newblock {\em SIGPLAN Not.}, 51(4):637--649, March 2016.

\bibitem{Grawinkel2015}
Matthias Grawinkel, Lars Nagel, Markus M\"{a}sker, Federico Padua, Andr{\'e}
  Brinkmann, and Lennart Sorth.
\newblock Analysis of the {ECMWF} storage landscape.
\newblock In {\em Proceedings of the 13th USENIX Conference on File and Storage
  Technologies}, FAST'15, pages 15--27, Berkeley, CA, USA, 2015. USENIX
  Association.

\bibitem{2017:SOSP:Canopy:Facebook}
Jonathan Kaldor, Jonathan Mace, Micha\l Bejda, Edison Gao, Wiktor Kuropatwa,
  Joe O'Neill, Kian~Win Ong, Bill Schaller, Pingjia Shan, Brendan Viscomi,
  Vinod Venkataraman, Kaushik Veeraraghavan, and Yee~Jiun Song.
\newblock Canopy: An end-to-end performance tracing and analysis system.
\newblock In {\em Proceedings of the 26th Symposium on Operating Systems
  Principles}, SOSP '17, pages 34--50, New York, NY, USA, 2017. ACM.

\bibitem{2016:D:SDA:Kiemle}
S.~{Kiemle}, K.~{Molch}, S.~{Schropp}, N.~{Weiland}, and E.~{Mikusch}.
\newblock {B}ig {D}ata {M}anagement in {E}arth {O}bservation: {T}he {G}erman
  satellite data archive at the {G}erman {A}erospace {C}enter.
\newblock {\em IEEE Geoscience and Remote Sensing Magazine}, 4(3):51--58, Sep.
  2016.

\bibitem{horison-storage-outlook}
Fred Moore.
\newblock {S}torage {O}utlook 2016.
\newblock \url{https://horison.com/publications/storage-outlook-2016}, 2016.

\bibitem{Murray_2017}
S~Murray, V~Bahyl, G~Cancio, E~Cano, V~Kotlyar, D~F Kruse, and J~Leduc.
\newblock An efficient, modular and simple tape archiving solution for {LHC}
  {R}un-3.
\newblock {\em Journal of Physics: Conference Series}, 898:062013, October
  2017.

\bibitem{2019:BiDS:Schindler}
Sirko Schindler, Marcus Paradies, and Andr\'e Twele.
\newblock {Here is my Query, where are my Results? A Search Log Analysis of The
  EOWEB\textsuperscript{\textregistered} Geoportal}.
\newblock In {\em 2019 Conference on Big Data from Space: Turning Data into
  Insights, (BiDS'19), Munich, Germany, 19-21 February, 2019.}, pages 1--4,
  2019.

\bibitem{VanDiepen15}
G.N.J. van Diepen.
\newblock {Casacore Table Data System and its use in the MeasurementSet}.
\newblock {\em Astronomy and Computing}, 12:174 -- 180, 2015.

\bibitem{Yan2017}
Wenrui Yan, Jie Yao, Qiang Cao, Changsheng Xie, and Hong Jiang.
\newblock {ROS: A Rack-based Optical Storage System with Inline Accessibility
  for Long-Term Data Preservation}.
\newblock In {\em Proceedings of the Twelfth European Conference on Computer
  Systems}, EuroSys '17, pages 161--174, New York, NY, USA, 2017. ACM.

\end{thebibliography}

\end{document}